\documentclass[conference,a4paper]{APSIPA2025}

\usepackage{amsmath,amsfonts}
\usepackage{graphicx}
\usepackage{multirow}
\usepackage{threeparttable}
\usepackage{booktabs}
\usepackage{comment}
\usepackage{xcolor}     
\usepackage{url}        
\usepackage[numbers,sort&compress]{natbib}  

\newcommand\blfootnote[1]{%
\begingroup
\renewcommand\thefootnote{}\footnote{#1}%
\addtocounter{footnote}{-1}%
\endgroup
}

\usepackage{geometry}
\usepackage{fancyhdr}

\fancypagestyle{firststyle}{
  \fancyhf{}
  \fancyhead[C]{}
}

\begin{document}

\title{MDD: a Mask Diffusion Detector to Protect Speaker Verification Systems from Adversarial Perturbations}

\author{
\authorblockN{
Yibo Bai\authorrefmark{1}, Sizhou Chen\authorrefmark{2}, Michele Panariello\authorrefmark{1}, Xiao-Lei Zhang\authorrefmark{3} \authorrefmark{4}, Massimiliano Todisco\authorrefmark{1} and Nicholas Evans\authorrefmark{1}
}

\authorblockA{
\authorrefmark{1}
EURECOM, France \\
E-mail: \{bai, panariel, todisco, evans\}@eurecom.fr}

\authorblockA{
\authorrefmark{2}
The University of Sydney, Australia \\
E-mail: szchen1005@gmail.com}

\authorblockA{
\authorrefmark{3}
Northwestern Polytechnical University, China \\
E-mail: xiaolei.zhang@nwpu.edu.cn}

\authorblockA{
\authorrefmark{4}
Institute of Artificial Intelligence (TeleAI), China Telecom Corp Ltd }
}

\maketitle
\thispagestyle{firststyle}

\blfootnote{This work is supported with funding received from the French Agence Nationale de la Recherche (ANR) via the P-SPIKE (ANR-23-CE39-0005).}

\pagestyle{fancy}
\begin{abstract}
Speaker verification systems are increasingly deployed in security-sensitive applications but remain highly vulnerable to adversarial perturbations. In this work, we propose the Mask Diffusion Detector (MDD), a novel adversarial detection and purification framework based on a \textit{text-conditioned masked diffusion model}. During training, MDD applies partial masking to Mel-spectrograms and progressively adds noise through a forward diffusion process, simulating the degradation of clean speech features. A reverse process then reconstructs the clean representation conditioned on the input transcription.
Unlike prior approaches, MDD does not require adversarial examples or large-scale pretraining. Experimental results show that MDD achieves strong adversarial detection performance and outperforms prior state-of-the-art methods, including both diffusion-based and neural codec-based approaches. Furthermore, MDD effectively purifies adversarially-manipulated speech, restoring speaker verification performance to levels close to those observed under clean conditions. These findings demonstrate the potential of diffusion-based masking strategies for secure and reliable speaker verification systems.
\end{abstract}

\section{Introduction}
Automatic Speaker Verification (ASV) plays a key role in providing secure access control to services, smart phones and devices. 
However, ASV systems are vulnerable to adversarial attacks~\cite{malacopula2024, szegedy2013intriguing, villalba2020x} whereby subtle and even imperceptible perturbations are added to an acoustic input to manipulate normal system behaviour, i.e.\ to obtain unauthorised access to protected services or devices.   
Such vulnerabilities pose a challenge to verification reliability~\cite{todisco2019asvspoof,lan2022adversarial, tan2022adversarial}. There is hence an interest to develop robust detection methods to protect against the threat of adversarial attacks.

Current research in defences against adversarial attacks falls into two categories~\cite{lan2022adversarial, tan2022adversarial}. Proactive defences, such as adversarial training, requires advance knowledge of specific attacks and continual retraining/adaptation of the ASV model to new attacks~\cite{lan2022adversarial}. Passive defences, in contrast, can be used to detect or eliminate adversarial perturbations without retraining. Among these, plug-in detection-based methods have received considerable attention due to their convenience and flexibility~\cite{lan2022adversarial}.

Various such detection schemes have been proposed recently. These include methods which rely on statistical analysis~\cite{hassan2024review,wu2024scalable}, auxiliary classifiers like learnable mask networks~\cite{chen2023lmd}, approaches based on self-supervised learning~\cite{wu2021improving}, and neural codec-based audio reconstruction~\cite{chen2024neural}. However, these approaches often face limitations. For instance, some are too computationally intensive for real-time applications or require large-scale pretraining, while others are dependent on prior knowledge of specific attack types. This reliance can leave such systems vulnerable to new, unknown attacks~\cite{li20r_interspeech, joshi22b_interspeech}.  
We propose a novel diffusion-based adversarial attack detector: the Mask Diffusion Detector (MDD). 
Our approach builds upon prior work in diffusion-based audio processing~\cite{bai2024diffusion,chen2024textual, bai2024adversarial}. However, whereas the forward process of traditional diffusion models typically acts to progressively reduce the different between the input and Gaussian noise ~\cite{ho2020denoising, sohl2015deep}, that of MDD is used to progressively produce a noised masked Mel-spectrogram. An 
embedding of the transcription text is applied during the conditional diffusion process to help preserve key information in the reconstruction.
In the reverse denoising and reconstruction phase,  the text-conditioned diffusion model learns to recover an estimate of the original, clean spectrogram. As a result, it can effectively identify and mitigate malicious, adversarial perturbations. Notably, MDD can be trained on bona fide data only, without any adversarial data. This attacker-independent approach promotes generalisation.
Compared to existing works, the main contributions are as follows. 
\begin{itemize}
    \item We introduce MDD, a novel adversarial attack detection framework built upon a text-conditioned masked diffusion model. By applying spectral masking in both the forward and reverse diffusion processes, MDD effectively adapts diffusion-based generative modeling to the task of adversarial detection in speaker verification.
    \item MDD requires neither adversarial training data nor large-scale pretraining, yet it achieves strong detection performance, demonstrating robustness and generalisation across attack types.
    \item Beyond adversarial detection, MDD preserves the performance of speaker verification (ASV) systems on clean data, ensuring practical applicability in real-world scenarios without compromising verification accuracy.
\end{itemize}

\section{Related work}

\subsection{Automatic speaker verification}
 
A typical ASV system consists of two stages: speaker embedding extraction and speaker similarity computation. Embeddings are extracted from an enrolment utterance $x_e$ and a test utterance $x_t$.
The similarity between the pair of embeddings is then computed to determine whether the speaker in $x_e$ matches that in $x_t$. Common speaker embedding extractors include x-vector~\cite{snyder2018x}, ECAPA-TDNN \cite{desplanques2020ecapa} and self-supervised learning-based models \cite{chen2022large}. 

\subsection{Adversarial attacks}

Given a bona fide (clean) utterance $x_t$, an attacker generates an adversarially perturbed version $x_{adv}$ which causes the ASV system to verify incorrectly the similarity between $x_{e}$ and $x_{adv}$. The adversarial example $x_{adv}$ is constrained to be perceptually similar to $x_t$ so that $\left\Vert x_{adv} - x_t \right\Vert_p \leq \varepsilon$, where $|\cdot|_p$ denotes the $\ell_p$-norm and $\varepsilon$ is a small positive constant. Common attack methods include the basic iterative method (BIM)~\cite{kurakin2018adversarial}, projected gradient descent (PGD)~\cite{madry2018towards}, and the fast gradient sign method (FGSM)~\cite{goodfellow2014explaining}.

\subsection{Diffusion models}

Diffusion models have emerged recently as powerful generative frameworks in various domains. They operate by adding noise gradually to input data and then by learning a de-noising process so that they can generate high-quality, clean data from pure noise inputs. By approximating source distributions through such an iterative process, diffusion models have been shown to perform well for generative tasks~\cite{rombach2022high, liu2023audioldm} and enhancement tasks~\cite{li2023diffusion, lu2022conditional}.

\section{Mask diffusion detector}
\begin{figure*}
    \centering
    \includegraphics[width=\linewidth]{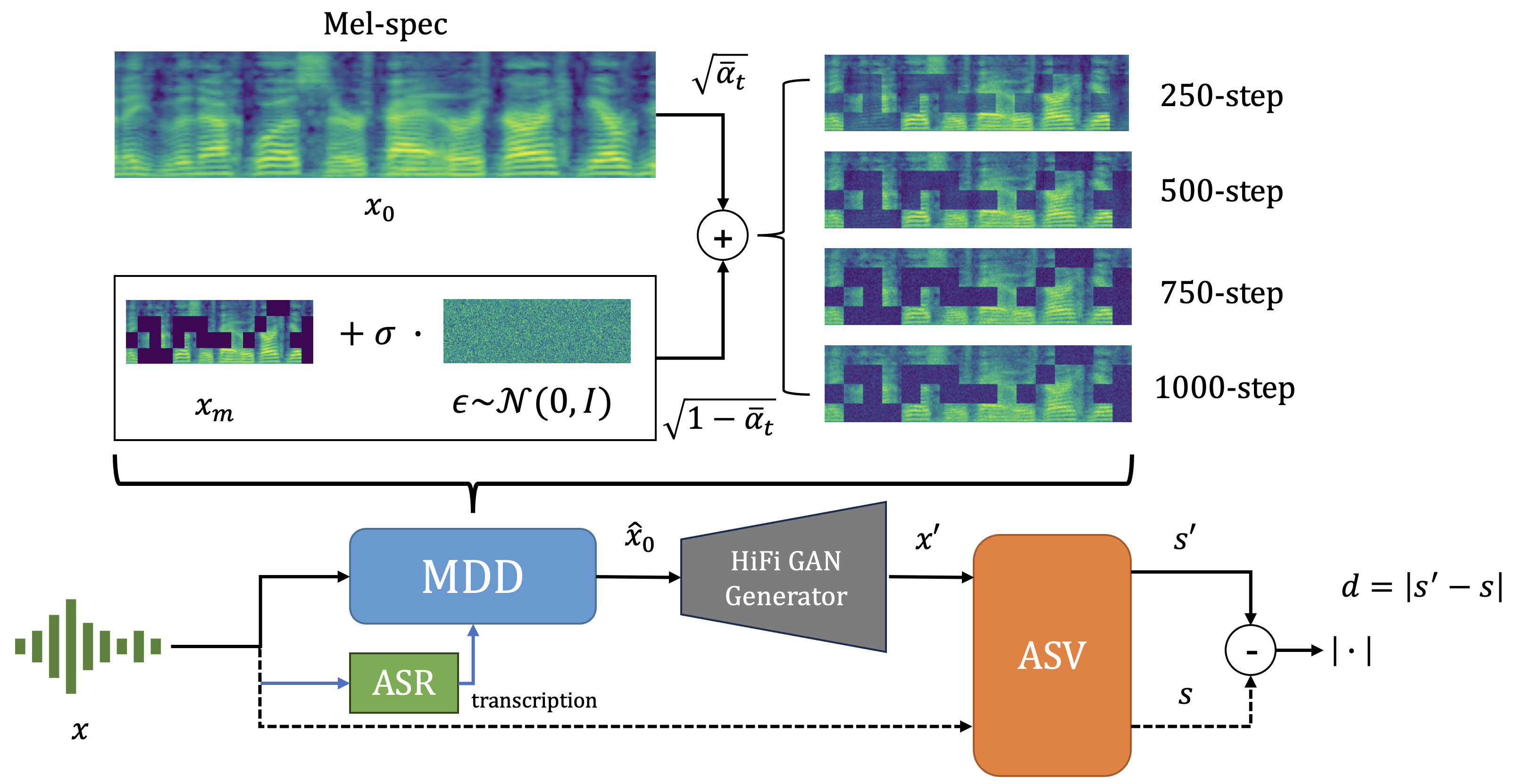}
    \caption{An illustration of the workflow of the proposed MDD method. Given an input Mel-spectrogram, MDD first generates a masked version of the feature and mixes it with Gaussian noise. During the forward process, the mask diffusion model applies masking and noise to the input step by step. In the reverse process, it performs unmasking and denoising operations conditioned with the transcription. A HiFi-GAN vocoder then reconstructs the waveform from the processed feature. Finally, both the original and the regenerated waveforms are fed into the ASV system to compute the score difference. }
    \label{MDD}
\end{figure*}

In this section we describe  the proposed Mask Diffusion Detector (MDD), a novel framework designed for the detection of adversarial perturbations aimed at deceiving ASV systems. MDD consists of two main components: a diffusion model adapted with a specific mask strategy and noise schedule, and a back-end detector to distinguish bona fide from adversarially-manipulated speech data. 

\subsection{Diffusion model with mask strategy}
An illustration of the MDD is shown in Fig.~\ref{MDD}. At the core is a diffusion model which iteratively denoises the masked Mel-spectrogram. The model encompasses a forward diffusion process and a reverse reconstruction process.  

\subsubsection{Forward diffusion process}
The forward diffusion process, denoted by $q(x_1, \dots, x_T | x_0, c)$, takes an initial clean Mel-spectrogram $x_0$ and a text condition $c$, and progressively adds noise and applies masking over a sequence of $T$  steps to produce $x_T$. 

First, a masked version of the input, denoted as $x_{m}$, is generated from the original, clean spectrogram $x_0$. This is achieved through random masking of $16 \times 16$ patch regions.

Second, a composite noise target, $N_{target}$, is formulated according to:

\begin{equation}
N_{target} = x_{m} + \sigma \cdot \epsilon ,
\label{eq:n_target}
\end{equation}
where $\epsilon$ is random noise sampled from a standard Gaussian distribution $\mathcal{N}(0, I)$ and with the same dimensions as $x_m
$, and $\sigma$ is a scalar coefficient used to control the signal-to-noise ratio in $N_{target}$.

The noisy spectrogram $x_t$ for a given timestep $t$ is subsequently generated according to a pre-defined noise schedule. It is generated from the initial clean spectrogram $x_0$, the new noise target $N_{target}$, and a noise schedule parameter $\bar{\alpha}_t$, which represents the cumulative product of noise variances up to step~$t$. In our modified framework, the standard Gaussian noise term is replaced by our specifically formulated $N_{target}$. Thus, $x_t$ can be expressed by the relation:
\begin{equation}
x_t = \sqrt{\bar{\alpha}_t} x_0 + \sqrt{1-\bar{\alpha}_t} N_{target}.
\label{eq:xt_definition}
\end{equation}
The forward diffusion hence steers $x_0$ towards a state which is a controlled combination of $x_0$ itself and the composite noise target $N_{target}$. Consequently, the masked input $x_m$ becomes an integral part of the degradation process which the reverse reconstruction process must learn to invert.

\subsubsection{Reverse reconstruction process}
The reverse process, $p_{\theta}(x_{t-1} | x_t, c)$, aims to reconstruct the initial clean Mel-spectrogram $x_0$ from corrupt version $x_t$, with parameter $\theta$ and a text condition $c$. Given that the forward process uses $N_{target}$ to noise and mask $x_0$, the model learns to denoise $x_t$ as well as to fill the masked area in $x_m$. At each step, the training objective is to minimise the reconstruction error between the predicted noise and $N_{target}$. By learning the distribution of clean data, MDD is able to remove adversarial perturbations from masked inputs.

\subsection{Waveform reconstruction}
Following the reverse diffusion process, the denoised Mel-spectrogram $\hat{x}_0$ is transformed back into a time-domain waveform using a pretrained HiFi-GAN vocoder~\cite{kong2020hifi}. HiFi-GAN is a non-autoregressive, GAN-based neural vocoder that maps Mel-spectrogram features to high-fidelity audio waveforms. In our framework, the vocoder operates as a fixed post-processing module, independent of the diffusion model training. Its purpose is to synthesise speech from the reconstructed spectrogram $\hat{x}_0$ such that it can be evaluated by downstream automatic speaker verification (ASV) systems. 

\subsection{Back-end adversarial detector}
A straightforward back-end detector is used to determine whether an input audio sample is clean or whether it contains adversarial perturbations.
The detector takes the form of a conventional ASV system which is applied separately to input signals $x$ and diffusion-purified version $x'$, thereby producing a pair of ASV scores $s$ and $s'$ respectively. The absolute score difference $d = |s - s'|$ is then calculated and thresholded for classification. 
For clean inputs, purification should produce $x'\approx x$ and $s' \approx s$, hence lower values of $d$.
In the case of adversarial inputs, then purification should remove the perturbations which would otherwise act to compromise the ASV system.  As a result, $s'$ (purified of perturbations) should be lower than $s$ (with perturbations) corresponding to comparatively higher values of $d$.  

An input sample is classified as adversarial if $d$ exceeds an empirically optimised detection threshold $\tau$. We set $\tau_{det}$ from experiments involving a set \( \mathbb{T}\) of clean data only (no adversarial examples) to achieve an arbitrarily-set target false positive rate \( FPR_\text{target} \) as follows:

\begin{equation}
    \tau_{det} = \min_{\tau} \left\{  \frac{|\{ x^i \in \mathbb{T} \mid d^i > \tau \}|}{\text{number of samples  in} \ \mathbb{T}} \leq FPR_\text{target} \right\},
\end{equation}
where $\tau \in \mathbb{R} $ and \( d^i \) denotes the score difference for the \( i \)-th clean sample.

\section{Experiments}

\subsection{Experimental setup}

We used the PGD method~\cite{madry2018towards} and a subset of 1,000 clean utterances extracted from the VoxCeleb1 test dataset \cite{nagrani2017voxceleb} to generate 1,000 adversarial examples with which to test the reliability of ECAPA-TDNN ASV and MDD systems in the white-box attack scenario. 

The input to the ECAPA-TDNN system is 80-dimensional log filterbank (LogFBank) components extracted with a 25 ms Hamming window and 10 ms frame shift. The model is trained using the VoxCeleb1 development set with the standard 512-channel architecture provided with the \textit{Wespeaker} toolkit~\cite{wang2023wespeaker}. The PGD algorithm is applied with 50 attack iterations and $\ell_2$-norm. 

We evaluate detection performance using the Detection Rate (DR) metric, which quantifies the percentage of adversarial examples correctly identified by the system. Specifically, DR is measured at fixed false positive rate (FPR) thresholds, ensuring a controlled trade-off between detecting attacks and minimising false alarms on bona fide inputs. A higher DR indicates a stronger capability to detect adversarial perturbations without significantly impacting clean audio samples.

We trained six MDD models with 0\%, 10\%, 25\%, 50\%, 75\% and 100\% masked Mel-spectrograms. 
Each model is trained for 10,000 iterations with a batch size of 4 using the LibriSpeech train-clean-100 subset. In MDD, the noise control factor $\sigma$ is set to 0.1, which we found to provide a good balance between mitigating adversarial effects and preserving the perceptual fidelity of the output audio. Our implementation is based on the \textit{audio-diffusion} toolkit\footnote{\url{https://github.com/teticio/audio-diffusion}}, and follows a default 1000-step DDPM noise schedule \cite{ho2020denoising}. We use the \textit{whisper-small} Automatic Speech Recognition (ASR) model\footnote{\url{https://huggingface.co/openai/whisper-small}} for transcription during conditional generation,
and \textit{Stella}\footnote{\url{https://huggingface.co/NovaSearch/stella_en_400M_v5}} to encode the text as conditional embeddings. A pretrained HiFi-GAN model \cite{kong2020hifi} from AudioLDM\footnote{\url{https://huggingface.co/cvssp/audioldm/tree/main/vocoder}} \cite{liu2023audioldm} serves as the vocoder in MDD. 

\subsection{Experimental results}
The DR results reported in Table~\ref{tab:dm} show that the 10\% masking configuration achieves the best detection performance across both FPR thresholds. Interestingly, the 0\% masking case (i.e., unmasked input) performs slightly worse, highlighting the effectiveness of introducing partial spectral masking during diffusion. As the masking ratio increases beyond 10\%, detection performance progressively declines due to greater information loss. These results suggest that moderate masking encourages the model to focus on key spectral regions relevant for detecting adversarial perturbations, while excessive masking degrades the model’s ability to reconstruct meaningful features.

\begin{table}[ht]
\centering
\caption{DR (\%) results for the MDD defence method at different mask ratios.}
\begin{tabular}{lcc}
\toprule
Mask ratio & FPR=0.1 & FPR=0.05 \\
\midrule
0\% (unmasked)         & 96.2 & 95.0 \\
10\%                   & \textbf{98.0} & \textbf{96.9} \\
25\%                   & 96.0 & 94.0 \\
50\%                   & 80.4 & 76.3 \\
75\%                   & 60.6 & 57.3 \\
100\% (fully masked)   & 55.4 & 47.6 \\
\bottomrule
\end{tabular}
\label{tab:dm}
\end{table}

\subsection{Comparison with other defence methods}
We compare the adversarial detection performance of the proposed MDD with 10\% masking against several existing defence methods, including the DAP diffusion model~\cite{chen2024textual} and three neural codec-based approaches: AcademiCodec~\cite{yang2023hifi}, SpeechTokenizer~\cite{zhangspeechtokenizer}, and DAC~\cite{kumar2023high}. The corresponding Detection Rate (DR) results under fixed false positive rate thresholds (FPR = 0.1 and 0.05) are reported in Table~\ref{tab:dr}.

\begin{figure}
    \centering
    \includegraphics[width=\linewidth]{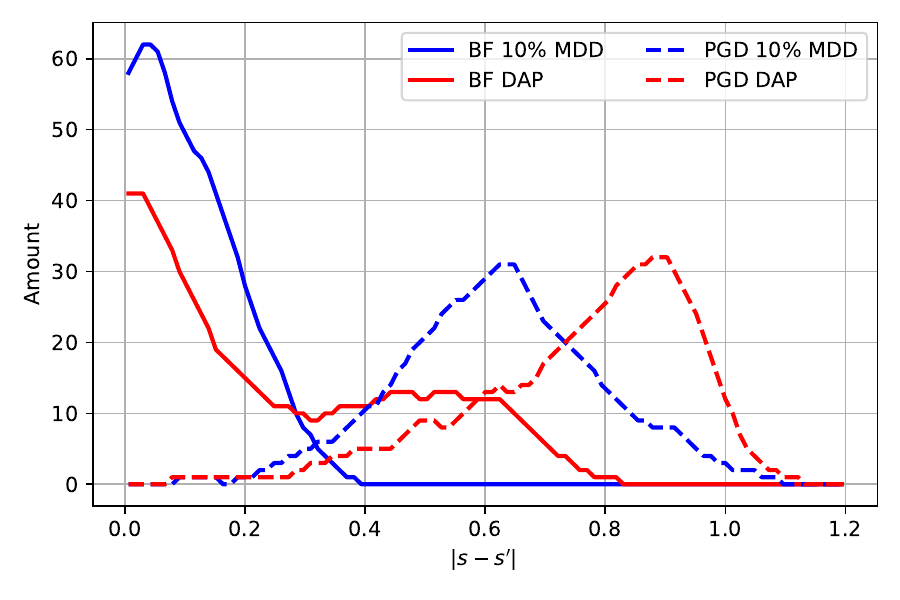}
    \caption{The distributions of score differences when applying 10\% MDD and DAP on bona fide (BF) and PGD adversarial data. A larger gap between the two lines of the same colour indicates that the PGD adversarial examples are more easily to be detected.}
    \label{fig:score distribution}
\end{figure}

The DAP method is based on a waveform-level diffusion model trained on unmasked audio using LibriSpeech, and fine-tuned on adversarial examples from the VoxCeleb1 development set. Although DAP achieves moderate DR values (78.0\% and 71.7\%), it falls short of the performance achieved by MDD, which attains 98.0\% and 96.9\% at the same FPR levels. Their score-difference distributions on bona fide and PGD adversarial data are shown in Fig. \ref{fig:score distribution}. This highlights the benefit of applying masking in the spectral domain and conditioning the reverse process with transcription in MDD. 

In contrast, the neural codec-based methods show substantially lower detection performance. While prior work has reported strong results using these codecs \cite{chen2024neural}, they typically rely on large-scale pretraining, and their performance drops significantly when retrained under the same conditions as MDD. Specifically, AcademiCodec, SpeechTokenizer, and DAC achieve DRs of 58.0\%, 65.7\%, and 74.7\% respectively at FPR=0.1, and even lower scores at FPR=0.05. These results suggest that neural codec-based detectors may lack generalizability and robustness when applied under constrained or mismatched training conditions.

Overall, the 10\% masked MDD demonstrates state-of-the-art performance among all compared methods, benefiting from its text-conditioned reconstruction and partial masking strategy, which collectively enhance its ability to suppress adversarial perturbations while maintaining the fidelity of clean speech.

\begin{table}[ht]
\centering
\caption{DR (\%) comparison between the 10\% mask MDD, DAP and neural codec-based methods under FPR=0.1 and FPR=0.05.}
\label{tab:codec}
\begin{tabular}{ll|cc}
\toprule
& Configuration & FPR=0.1 & FPR=0.05 \\
\midrule 
   MDD & 10\% Mask & \textbf{98.0}  & \textbf{96.9} \\
   DAP \cite{bai2024diffusion} & -  & 78.0 & 71.7 \\
   AcademiCodec \cite{yang2023hifi} & 16k-320d-l-uni &  58.0&  45.4\\
   SpeechTokenizer \cite{zhangspeechtokenizer} & hubert\_avg &  65.7&  59.9\\
   DAC \cite{kumar2023high} & 16k & 74.7 & 65.8 \\
\bottomrule
\end{tabular}
\label{tab:dr}
\end{table}

\section{Purification Impact on ASV Performance}

While detection accuracy is a critical metric for evaluating adversarial defences, the ultimate objective of any defence method is to protect the downstream application, which is ASV in our case. A truly effective purification system must not only detect and counteract adversarial perturbations, but also preserve ASV performance under real-world conditions.

To evaluate this, we assess the impact of different purification methods on ASV performance using the Equal Error Rate (EER) metric. Table~\ref{tab:adv_puri} reports EERs for both bona fide and PGD adversarial trials. We distinguish between two types of trials: (i) \textit{target vs. non-target}, which reflects performance for clean data, and (ii) \textit{target vs. adversarial non-target}, which evaluates robustness under attack.
Without purification, the ASV system achieves an EER of 1.4\% for bona fide trials but fails dramatically under adversarial conditions, reaching 73.2\%. This result highlights the vulnerability of unprotected ASV pipelines.

In contrast, MDD significantly reduces the EER for adversarial inputs, especially for masking ratios of 10\% and 25\%, which achieve 18.0\% and 17.6\% respectively, more than a fourfold reduction compared to the unprotected baseline. Importantly, the 10\% MDD model also maintains a low EER (4.0\%) for clean data, striking the best balance between robustness and reliability.
Higher masking ratios (50\% and above) lead to performance degradation for bona fide data due to excessive information loss during reconstruction. This underscores the importance of using a moderate masking strategy to retain speaker-discriminative features while suppressing adversarial noise.

Compared to other approaches, including the diffusion-based DAP model and several neural codec-based methods (AcademiCodec, SpeechTokenizer, DAC), MDD consistently achieves superior EERs across both clean and adversarial conditions. These alternative methods suffer from poorer generalisation and higher EERs, even when retrained under the same conditions.
In summary, MDD provides not only strong adversarial detection, but also practical and effective purification for robust ASV, demonstrating its potential as a viable defence mechanism for real-world speaker verification systems.

\begin{table}[ht]
\centering
\caption{Purified ASV EER (\%) results on bona fide data and PGD adversarial data with different methods. "tar" refers to target speaker trials, "non-tar" refers to non-target speaker trials, and "adv" refers to adversarial non-target trials.}
\begin{tabular}{lcc}
\toprule
\textbf{Purification method} & \textbf{Bona Fide EER (\%)} & \textbf{PGD EER (\%)} \\
 & \textbf{tar vs. non-tar} & \textbf{tar vs. adv} \\
\midrule
No Purification    & 1.4  & 73.2  \\
\midrule
0\% MDD (unmasked)       & 6.0  & 19.2  \\
10\% MDD           & \textbf{4.0}  & 18.0  \\
25\% MDD           & 6.2  & \textbf{17.6}  \\
50\% MDD           & 18.8 & 27.6  \\
75\% MDD           & 40.6 & 43.4 \\
100\% MDD (fully masked) & 49.4 & 50.0 \\
\midrule
DAP \cite{bai2024diffusion}               & 31.2 & 31.4  \\
AcademiCodec \cite{yang2023hifi}       & 51.4  & 51.4  \\
SpeechTokenizer \cite{zhangspeechtokenizer}    & 40.6  & 40.8  \\
DAC \cite{kumar2023high}              & 33.0  & 35.0  \\
\bottomrule
\end{tabular}
\label{tab:adv_puri}
\end{table}

\section{Conclusions}

In this work, we introduced MDD, a novel adversarial defense framework designed to protect automatic speaker verification (ASV) systems against imperceptible perturbations. MDD leverages a \textit{text-conditioned masked diffusion model}, which progressively denoises masked Mel-spectrograms while preserving essential speaker information through transcription-based conditioning.

Unlike many existing approaches, MDD does not rely on adversarial training or large-scale pretraining, yet it achieves strong performance in both detection and purification tasks. Our experiments demonstrate that a moderate masking ratio (specifically 10\%) yields the best trade-off, allowing MDD to effectively identify adversarial examples while maintaining high verification accuracy on clean speech.

We performed comprehensive comparisons with state-of-the-art diffusion-based and neural codec-based purification methods. MDD consistently outperforms these baselines in terms of detection rate and purified ASV equal error rate (EER), confirming its robustness and practical applicability.

Importantly, we emphasise that the ultimate goal of adversarial defence is not only to detect attacks, but to ensure the reliability of ASV systems in real-world deployment scenarios. Our results show that MDD meets this goal, providing a lightweight, effective, and generalisable solution for securing voice-based authentication systems.

\bibliographystyle{IEEEtran}
\bibliography{ref}
\end{document}